# Aging vs crystallisation dynamics in hyperquenched glasses and a resolution of the water $T_g$ controversy.

Y-Z. Yue* and C. A. Angell#

* Danish Center for Materials Technology, Aalborg University, 9220 Aalborg, Denmark

#Dept. of Chemistry and Biochemistry, Arizona State Univ. Tempe, AZ85287

**The formation of glasses on cooling is normal for substances of wide liquid range (large $T_b/T_g$ >2.0 for molecular liquids) and can be induced for most liquids if cooling is fast enough. The former group of liquids exhibit glass transformations (abrupt jumps of heat capacity as glass becomes supercooled liquid) during reheating at normal scanning calorimetry rates (~1K/s), whereas the latter usually pass directly to the crystalline state. Water (vitrified by hyperquenching or vapor deposition) is an intermediate case and the possibility of observing a glass transition before crystallisation has been debated vigorously but inconclusively over five decades [1,2]. The consensus for the last two decades has been to accept a glass transition at 136K [3,2], but this transition has perplexing qualities [4]. Recently it has been suggested, using enthalpy relaxation arguments [2,5], that this assignment must be wrong. The re-assignment of $T_g$ to temperatures above the crystallisation temperature has been vigorously contested [6]. Here we use detailed anneal-and-scan studies of a hyperquenched inorganic glass, which does not crystallize on heating, to interpret the perplexing aspects of the 136K phenomenon in water. We** **show that it is indeed linked to a glass transition, though only via a cross-over phenomenon. The thermal history that gives the same behaviour ("shadow" glass transition) in the inorganic glass is linked by crossover to a "normal" glass transition 23% higher in temperature. While this confirms that a normal glass transition is indeed unobservable for water, the vitreous nature of hyperquenched glassy water is strongly supported. The shadow glass transition is reproducible in the inorganic glass as it is in $H_2O$. The observed aging dynamics are very relevant to current glass theories, in particular to the interpretation of dynamical heterogeneity which is seen to have an energy manifestation. (270 words)**

The most abundant form of water in the universe is the glassy form, which exists in thin condensed films on interstellar dust particles [7]. This glass is in a configurationally excited, or high fictive temperature state, relative to glasses formed by normal cooling processes, and this circumstance has lead to much confusion in the interpretation of its nature. Studies of this highly unstable state are usually made by heating the glass in controlled fashion, thereby allowing it to relax out of the high energy state, but invariably leading it to crystallise in the range 150-160K. There is no question that the deposited or hyperquenched material is able to relax. The source of dispute concerns the state to which it is able to relax before the crystallisation occurs.

Until recently it was agreed that it reaches the internally equilibrated (or metastable liquid) state. It was believed to reach this state via a broad glass transformation (between 136 and 150K when scanned at 10K/min), and to be able to exist in this "ultraviscous state"[8] for considerable intervals of time. Indeed, it seemed that by appropriate application of pressure it could be cycled between this LDA (low density amorphous) water form and a high density amorphous HDA form, repeatedly [9]. Such a view is supported by the observation of various relaxation processes, not only between polyamorphic states, but internally within one form according to a variety of spectroscopic features [10] and also according to "blunt-probe" penetrometry [11].

What has changed this perception is the realization that hyperquenched glassy states of systems that are *not* abruptly crystallising during reheating, exhibit features that are inconsistent with the behaviour of water under the assumption that it has a glass transition at 136K [5]. In the present work we take the opposite approach and reach the same conclusion. We use a non-crystallising hyperquenched glass, that is readily available, to make detailed time-temperature studies of the annealing and recovery processes that occur when the hyperquenched glass is given the chance to relax. We then identify protocols that reproduce phenomena that share all the perplexing features of the 136K vitreous water phenomenon. However because the system under study is a good glassformer (doesn't crystallise) it offers the possibility of observing how the perplexing phenomena relate to the "normal" glass transition of this system. We have carried out this study on an inorganic glass because of its availability and simplicity in handling (the material is non-hygroscopic and woolly in texture [12]). However, these studies could equally well be performed on a molecular "good" glassformer vitrified by hyperquenching or vapour deposition. The surprise is only that it has not been performed long before. Results like the present ones have in fact very recently been obtained on electrospray-quenched dibutylphthalate glass (L.-M. Wang and C. A. Angell, to be published).

An advantage of the present study is that it not only assists us in understanding vitreous water to an unprecedented level, but it also provides valuable insights into the out-of-equilibrium glassy dynamics that is currently under intense study by theoretical physicists using model liquids studied by molecular dynamics [13]. In particular it emphasises the importance of the non-exponentiality and non-linearity of relaxation processes in glasses that is manifested in these studies in a particularly striking way.

Fig. 1 shows the relation between the enthalpy and the cooling rate according to a standard textbook figure (the right panel), and connects this diagram to the energy at which the glass is trapped for different severities of quenching. A trap is otherwise referred to as a "basin of attraction" on the multidimensional energy hypersurface [14] which is currently much used in discussion of the energetics and kinetics of glassformers [15,16]. Data to be shown below show how the common idea that a complex system can be described as a point moving on such a surface is in need of refinement.

To examine the manner in which a system, that has been trapped at abnormally high levels of such a landscape by fast quenching, recovers the state of a normal glass, we first define a "standard" glass. A standard glass is one that has been cooled into the glassy state at a rate $Q = -0.33K/s$ (20K/min), as represented by the lower curve of Fig 1. 0.33K/s is chosen because the "onset-heating glass temperature", $T_{g,onset}$, defined as in Fig.2 inset, and ref.2, from a differential scanning calorimetry (DSC) "standard scan", is then the temperature at which the structural relaxation time is ~100s [17]. The standard scan is a DSC upscan conducted at the same rate, Q = +0.33K/s. Then we compare, with this standard glass upscan, the 0.33K/s DSC upscan of a hyperquenched glass, which is a glass formed by cooling into the glassy state along a curve like curve 3 of Fig.1, Q = $10^6$ K/s.

For this study we chose a naturally occurring silicate glassformer (see Methods section) that has been transformed into a 2-7μm fiber wool by a centrifugal melt spinning process [12]. Carried out at a temperature of ~1800K, this provides a quenching rate estimated at $10^6$ K/s [12], comparable to that used in the vitrification of water [3,18]. The solid curve in Fig.1 inset is the initial scan of this material and the dashed curve is the rescan of the same sample after cooling back into the glassy state at a "standard rate" of 0.33K/s (20K/min). A continuation of either scan to higher temperatures results in a crystallization at 1156K.

The state of the hyperquenched glass can then be described by its excess enthalpy, which is given by the area between hyperquenched and standard scans, or by its "fictive" temperature. The fictive temperature is indicated in Fig. 1 by the breakaway point from the equilibrium liquid enthalpy, and it of course depends on the quenching rate. The manner in which the fictive temperature (1155 K in the present case) is determined from data in Fig. 2 inset, has been described in detail elsewhere [5,12]

In order to understand why water was incorrectly assigned a $T_g$ of 136K on the basis of DSC studies of hyperquenched water, we must understand how the annealing process used by Johari et al [6] to reveal a glass-transition-like thermal event, affects the subsequent upscans of a non-crystallising hyperquenched glass. To overturn the earlier conclusions [19,20] that no $T_g$ can be measured for vitreous water Johari et al first annealed the hyperquenched samples for 90 min at 130K, and then rescanned them from low temperatures. Correct annealing can enhance the visibility of a glass transition, at the expense of increasing the $T_g$ by a few %. In Fig 3 we show a series of such anneal-and-scans for the present glass samples in which the annealing temperature has been varied from 0.5 to 0.9 $T_{g,onset}$. We deliberately stop short of the annealing range in which the overshoot characteristic of the glass transition is enhanced.

The significance of this series of scans to current debate in glass physics will be outlined below, after emphasising the resolution of the water $T_g$ conundrum that they permit. To explain this resolution we could select either of scans g or h of Fig. 3 for closer examination. However we can introduce extra information by choosing an equivalent scan obtained by annealing a sample longer at a lower temperature before scanning. Thus in Fig. 4 we compare a scan obtained by annealing one day at 773K (instead of 90 min at 823K) with the scans that have provided the basis for the 136K assignment of $T_g$ for water. The lower panel of Fig. 4 shows results for both hyperquenched (HQG) and vapor-deposited (ASW)[21]. Both materials crystallize at ~150K, and this is the source of the confusion, as we now see.

The temperature scales in Fig. 4 are chosen so that 0 K is at a common point (offscale). With this choice, features of the two parts of figure 4 may be compared on a common basis, notwithstanding their very different temperature ranges. It is immediately clear from this comparison that the feature assigned to the "$T_g$ of water" in ref. 3 and other papers [4,22] is, instead, a very interesting feature of the annealing kinetics of hyperquenched glasses. We will call this feature a 'shadow' glass transition because it is indeed an image of the "real" or standard onset $T_g$. The real $T_g$ however, occurs at a temperature some 20-25% higher in absolute temperature, according to Figs. 3 and 4. We note that Velikov et al [5] were incorrect in assigning the 136 K endotherm to a special phenomenon (Bjerrum defect relaxation) peculiar to water. The same effect will be seen in any glassformer, though the details may vary according to the distribution of relaxation times (non-exponentiality, and non-linearity of relaxation, characteristic of the glassformer. The important conclusion is that the endotherm previously attributed to the glass transition of water is an annealing pre-peak, not a normal glass transition. The latter remains unobservable, hidden by crystallisation as argued in refs. 5 and 20. The quenched-in enthalpy that is unrelaxed at the temperature of crystallisation of HQG, (which prompted the analysis of Velikov et al [5]), corresponds to the area lying *under* the standard scan in Fig. 4, and *above* the crossover temperature in the case of the non-crystallising glass.

The "annealing prepeak" at 710- 850K in the inorganic glass can be reproduced any number of times by repeating the annealing process at the same temperature, *provided that* the crossover temperature is not exceeded on heating [23]. Reproducibility of the endothermic effect was one of the supporting arguments that has previously lead to general acceptance of 136K as the $T_g$ appropriate to water. Of course avoiding the crossover was mandated by the need to avoid crystallization.

Before considering the significance, to current issues in glass physics, of this striking behaviour of the present system, let us quickly point out the explanation of the other puzzles that the present scenario permits.

The first puzzle is the very small heat capacity jump $\Delta C_p$ at $T_g$ according to Fig. 4, lower panel, acknowledged in ref. 21. It is tenfold smaller than that predicted by extrapolations of aqueous solution data [20,23], (which was taken as evidence for a "transition" in ref. 4). It is now explained by the comparison of $\Delta C_p$ values for "shadow" and "real" glass transitions of the present glass, as exhibited in Fig. 4. Secondly, the reason for the sudden jump (~25K) in $T_g$ of hyperquenched propylene glycol +water solutions, which occurs at the composition where the solutions become glassforming on normal cooling [22], becomes obvious. Thirdly, the reason that water appears to be such a strong liquid according to analysis of its transition width [2] is that the shadow transition is intrinsically broader than the real transition (by a factor of 2, in Fig.4). Since we now cannot measure water liquid state properties below 235K, except on nanosecond time scales during hyperquenching, the matter of its fragility near its $T_g$ becomes moot, unless Maxwell demons are introduced, or nanoscopic samples invoked, to eliminate crystallisation.. For a "Maxwell demon-protected" form of non-crystallising water, scaling by the ratio of shadow $T_g$ to standard $T_g$ of the non-crystallising glass suggests a "hidden $T_g$" for water of 160--180K, depending on how the effect of increasing network character near pure water is assessed.

What does it mean that the crystallisation of ice $I_c$ is so slow to occur, when the glass can relax and even flow [11] so far below the "real" glass temperature? The fact that this relaxation and even viscous flow can occur while parts of the structure carrying considerable stored energy are still quite unrelaxed (curve h other sources e.g. Fig 9 of ref. 24]), implies a very heterogeneous dynamics for the glass. The failure to nucleate, under unstable conditions (only diffusion is required) then suggests that the elements of the structure that are relaxing below the temperature $T_c$ in Fig. 4 (lower panel), are too small in length scale to support a critical nucleus.

This evidence for heterogeneity brings us to consider the broader implications of Figs. 3 and 4. We find the crossover from endothermic to exothermic behaviour (relative to that of the standard-glass-standard-scan) by the annealed hyperquenched "good" glassformer to be very significant. It may be interpreted as a strong indication that the source of the dynamic heterogeneity currently being found in so many glassformers [25-32] has a counterpart in the potential energy, hence in the structure. Heretofore it has been widely held that it is a purely dynamical effect, mainly because there was no evidence to the contrary. But it is much easier to detect energy changes than structural changes. Changes in potential energy can only come from changes in vibration averaged structure.

The striking aspect of curves f, g and h of Fig. 3 and the scan in Fig. 4, is that the slowly relaxing elements of the structure are of high energy, relative to the standard glass. They remember their quenched-in states longer than those elements which can relax reversibly at low temperatures. This low temperature relaxation can be repeated over and over while the high energy structure responsible for the exotherm at 900-950 K remains locked in place. The system can even undergo the macroscopic deformation necessary to define a steady state viscosity while this structure remains locked in [25]. Since this is the opposite of what is implied by such ensemble average concepts as are embodied in the Adam-Gibbs entropy theory [29], and in the Cohen and Turnbull free volume model [30], a regio-specific interpretation must be sought. One that presents itself naturally is that these high energy elements are slowly relaxing because they are characterised by large length scales that were frozen-in during the hyperquench. It is the length scale, not the energy, that is important to the relaxation, this suggests. Another view is that, in the high energy state frozen in by the hyperquench, a distribution of loose and tighter packings exists which is uniformly higher in energy than the corresponding distribution at lower fictive temperatures. Then the fast relaxing elements are indeed the higher energy elements and the slower components are only high in energy relative to the fast elements in the standard glass. Needed to choose between the two are comparative measurements of relaxation rates in the standard glass and the hyperquenched, and incompletely relaxed glass, at the same sub-$T_g$ temperature. Whether or not the hyperquenched glass structure can provide a matrix within which a permanent high mobility of a fraction of the molecules can exist, is the question of interest. This is the "vault effect" discussed by Donth [34], and the evidence from conductivity studies in ionic glasses is that it is real.

Where does this leave the energy landscape concept that has been receiving so much attention in recent work? In the landscape paradigm [31,14-16] the system is represented by a single point moving on the multidimensional potential energy "hypersurface" (i.e., landscape), in effect "hopping" from one minimum to another. A single point cannot easily be seen as simultaneously moving both towards and away from the "normal" energy for the glass (i.e. that cooled according to the lower curve in Fig.1) as our present system appears to be capable of doing. This suggests the need for some modification, (and complication), of the landscape concept to permit the total system to be replaced by an ensemble of dynamically independent systems of different sizes and associated relaxation times. But then much of the simplicity, hence appeal, of the landscape concept is lost. The reconciliation must lie, presumably, in the understanding of dynamics in higher dimensional spaces. The present observations may then be seen as offering some guidelines in the development of such dynamics.

In summary, we have identified the true origin of the endothermic effect observed at 136 K during DSC scanning of annealed amorphous water and previously attributed to its glass transition. It is an annealing prepeak of a form which, in hyperquenched glasses, occurs some 20% lower in temperature than the standard glass transition to which it is related. This confirms that glassy water remains glassy until it crystallizes but supports the notion that glass transition-like effects observed at ~160 – 180 K in hydrated proteins and hydrogels are due to water structure unfreezing.

# Methods section.

The fibre samples were spun from a basalt-like glass melt with the composition: 49.3 $SiO_2$, 15.6 $Al_2O_3$, 1.8 $TiO_2$, 11.7 FeO, 10.4 CaO, 6.6 MgO, 3.9 $Na_2O$, and 0.7 $K_2O$ (wt%) in the temperature range between 1473 to 1573 K. Before the spinning, the glass melt was homogenised at temperatures of about 1800 ~ 1850 K for five hours. The fibres were spun using the wheel centrifuge process, which has also been called the cascade or mechanical spinning process. The glass melt is conveyed down a trough and onto the outside of the rim of a spreader wheel. While some material is spun off, most of the melt is transferred to either one or two larger wheels, which are located close to the first wheel, and which are spinning in the opposite direction. Most of the fibres produced in this process are generated from these wheels. As droplets of the melt are thrown from the wheels by centrifugal force, a fibre is generated between the droplet and its starting point as they diverge from each other due to the swift rotation of the wheel. A portion of the melt is fiberised on each wheel. The fibres were drawn with speeds of over 200 m/s, so that the average diameter of the fibres could reach 3.5 μm and the average hyperquenching rate could reach $4\times10^6$ K/s.

To get rid of the "pearls" from the fibres and ensure uniform fictive temperature and properties, the fibres were sieved using a 65 μm sieve. The cooling rate of the fibres was calculated from the relation between cooling rate (Q) and shear viscosity (η): log ($1/q_c$) = log η - 11.22, as described in Ref. [37]. Before the DSC measurements, ageing of the fibres is performed in an annealing furnace at various temperatures for 90 minutes.

The heat capacity of the fibres was measured by using a differential scanning calorimeter (DSC) Netzsch STA 449C. The fibres were placed in a platinum crucible situated on a sample holder of the DSC at room temperature. To obtain the data in Figs. 2-4 samples were held 5 mins at an initial temperature of 333 K, and then heated at a "standard" rate of 0.333 K/s to the temperature 1013 K, and then cooled back to 573K at a rate of 0.333 K/min to 573 K, forming the "standard" glass. After natural cooling to room temperature, the second upscan was performed using the same procedure as for the first. To determine the heat capacity ($C_p$) of the fibres, both the baseline (blank) and the reference sample (Sapphire) were measured. In order to confirm reproducibility, the measurements for some samples were repeated to check whether drift in the baseline occurred or not, which affects the measurements. We have also measured samples of continuously drawn fibres, which have unique diameters. The results show the same relaxation behaviour as that of the fibre (wool) made by the cascade process, indicating that the narrow distribution of diameters of the sample studied in this work has negligible influence on the fictive temperatures and shapes of the $C_p$ curves.

## References.

# FIGURES FOR Yue and Angell

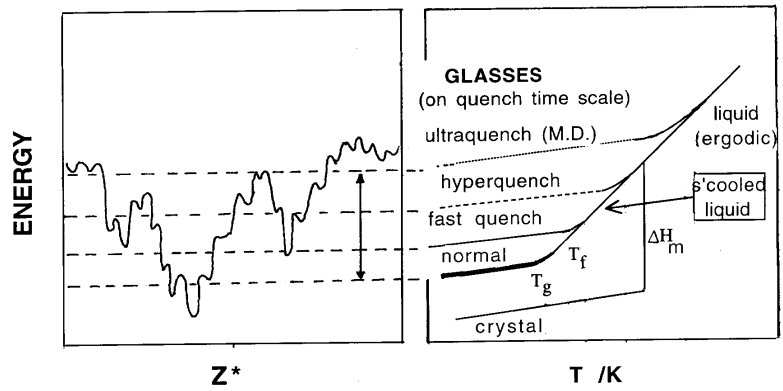

Fig. 1. Depiction of the commonly supposed relation between rate of the quench and the energy of the "landscape" basin of attraction in which the system is trapped. We show below how the depiction of the system as a point moving on such a surface is inadequate for systems accessible to experiment

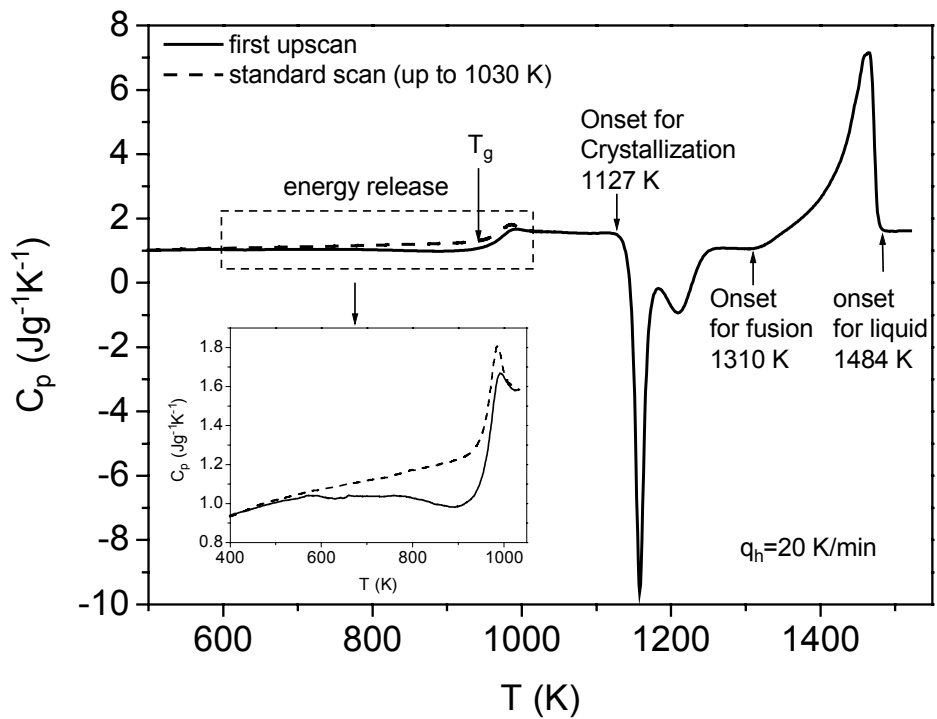

**Figure 2**. Complete DSC 20K/min upscan of melt-spun glass sample showing initial relaxation (weak on this scale) followed by glass transition near 960K, crystallization at 1156 K, and finally liquidus temperature at 1484K.

**Inset** shows superposition of the glass transition region of main figure with the "standard scan" of the sample *after* completion of an intial scan to 1030K which relaxes the quenched-in enthalpy without permitting any crystallization. The standard onset $T_g$ for this glass is 944K [21] The areabetween the curves is the energy difference between the hyperquenched and normal glass, as indicated by the vertical double arrow on the left panel of Fig.1 and can be used to determine the fictive temperature of the hyperquenched glass.

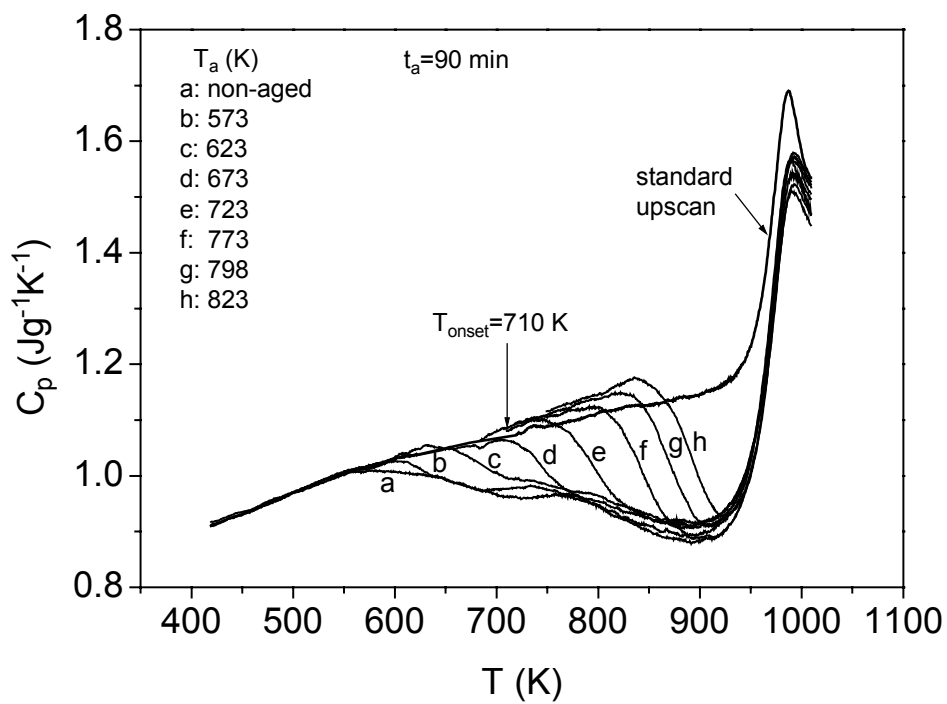

Figure. 3. Comparison of the two scans of Fig. 2 inset (here marked "standard" and "a") with a series of scans of aged samples. For each of these (b to h) a fresh sample of the hyperquenched glass was annealed 90 min. at one of the temperatures listed in the legend before lowering the temperature to 400K and scanning up to 1030K to observe the enthalpy relaxation of the remaining excess (frozen-in) enthalpy. Note, in particular, the evidence that the samples aged at temperatures above ~723K reach lower enthalpies than the standard glass for fast-relaxing components of the structure, while the slow-relaxing components of the structure are *still* trapped in states of much higher enthalpy than that of the standard glass. This proves that, if the non-exponential relaxation of viscous liquids is due to dynamical heterogeneity as is now suggested, then the heterogeneity is not merely dynamic but has an energetic, hence structural, origin.

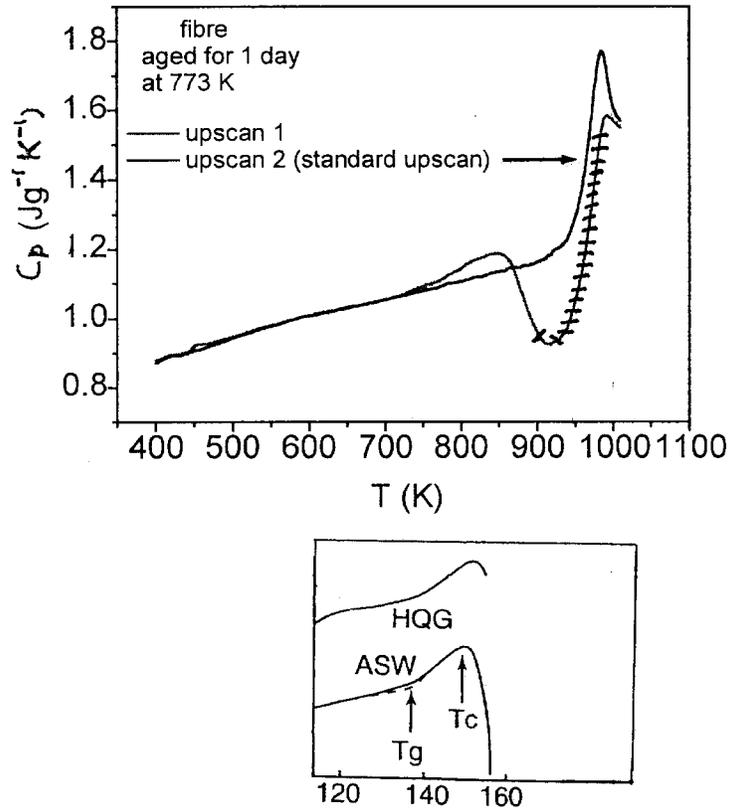

Figure 4. Comparison of the thermogram of the hyperquenched mineral glass sample aged one day at 773K before upscanning, (upper panel) with the upscans of hyperquenched water HQG and vapor deposited ASW samples [21] that had been aged 90 mins at 130K before upscanning (lower panel). The temperature scales are set so that each has 0K at the same point offscale., These latter have been the basis for assigning a $T_g$ of 136K to water. The comparison shows that this endotherm can arise from an annealing effect that we here call the "shadow glass transition" because it is indeed related to the glass transition. However, as is obvious from the comparison with the non-crystallizing glass, the "real" glass transition in the case of water has been eliminated from observation by crystallization. The hatched line in the upper part of the figure is the part that we suggest has been cut out of the water sample scan by crystallization commencing at or slightly above the temperature marked $T_c$. The ratio of the onset temperature of the shadow $T_g$ to the ("standard onset") $T_g$ of the mineral glass, is **0.80**. Applying the same ratio to the "shadow $T_g$" of water we obtain a "real but hidden" $T_g$ of **169K**. However, this ratio will depend on system fragility so the estimate of the hidden $T_g$ for water is uncertain. (Data and $T_g$ identification on HQG and ASW taken from Ref. 21.)